\documentclass[sigconf]{acmart}

\usepackage{algorithm}
\usepackage{algpseudocode}

\usepackage{microtype}

\usepackage{booktabs}
\usepackage{multirow} 
\usepackage{pifont}
\newcommand{\cmark}{\ding{51}}

\usepackage{listings}
\usepackage{xcolor}

\lstdefinelanguage{SPARQL}{
  morekeywords={SELECT, WHERE, SERVICE, BIND, FILTER, OPTIONAL, AS, PREFIX, DISTINCT, FROM, NAMED, GROUP, BY, ORDER, LIMIT, OFFSET, HAVING, UNION, MINUS, VALUES, IF, BOUND, STR, SUBSTR, STRLEN, CONTAINS, STRAFTER},
  sensitive=true,
  morecomment=[l]{\#},
  morestring=[b]",
  morestring=[b]',
}

\setcopyright{none}
\renewcommand\footnotetextcopyrightpermission[1]{}
\acmConference{}{}{}
\acmBooktitle{}
\acmDOI{}
\acmISBN{}
\begin{document}


\emergencystretch 3em

\title{Deontic Knowledge Graphs for Privacy Compliance in Multimodal Disaster Data Sharing}

\author{Kelvin Uzoma Echenim}
\affiliation{
  \institution{University of Maryland, Baltimore County}
  \city{Maryland}
  \country{USA}}
\email{kelvine1@umbc.edu}
\orcid{1234-5678-9012}

\author{Karuna Pande Joshi}
\affiliation{%
  \institution{University of Maryland, Baltimore County}
  \city{Maryland}
  \country{USA}}
\email{karuna.joshi@umbc.edu}
\orcid{1234-5678-9012}

\begin{abstract}
Disaster response requires sharing heterogeneous artifacts, from tabular assistance records to UAS imagery, under overlapping privacy mandates. Operational systems often reduce compliance to binary access control, which is brittle in time-critical workflows. We present a novel deontic knowledge graph-based framework that integrates a Disaster Management Knowledge Graph (DKG) with a Policy Knowledge Graph (PKG) derived from \texttt{IoT-Reg} and FEMA/DHS privacy drivers. Our release decision function supports three outcomes: \textsc{Allow}, \textsc{Block}, and \textsc{Allow-with-Transform}. The latter binds obligations to transforms and verifies post-transform compliance via provenance-linked derived artifacts; blocked requests are logged as semantic privacy incidents. Evaluation on a 5.1M-triple DKG with 316K images shows exact-match decision correctness, sub-second per-decision latency, and interactive query performance across both single-graph and federated workloads.\end{abstract}

\keywords{Privacy Compliance, Knowledge Graphs, Deontic Logic, Disaster Management, IoT, Semantic Web}

\maketitle

\section{Introduction}
In the immediate aftermath of a large-scale disaster, prompt situational awareness is the currency of effective response. Emergency management agencies increasingly rely on multimodal data ranging from Unmanned Aircraft Systems (UAS) imagery and remote sensing feeds to tabular survivor registries to assess damage, identify hazards, allocate resources, and coordinate search-and-rescue operations \cite{pak2025situational}. This data myriad facilitates a \textit{common operational picture} essential for interagency coordination. However, the mass collection of visual data in residential areas inevitably captures Personally Identifiable Information (PII), such as faces and license plates, raising significant privacy concerns \cite{disaster_privacy_2020, dhs2024aircraft}.

\subsection{Policy-Aware Disaster Data Sharing}
Federal agencies operate under strict regulatory mandates to safeguard this data. For instance, the Federal Emergency Management Agency (FEMA) in the United States explicitly states in its Privacy Impact Assessment (PIA) for UAS that while it does not target individuals, any \textit{incidental} PII collected must be obfuscated or deleted prior to storage or dissemination \cite{dhs2020uas_pia}. Similarly, FEMA's System of Records Notice (SORN DHS/FEMA-008) mandates strict access controls and audit logging for inter-agency PII transfers \cite{dhs2024sorn}. The urgency of disaster response often conflicts with the procedural rigidity of these compliance reviews \cite{disaster_privacy_2020}. Emergency managers face a critical dilemma: release data immediately to aid response and risk privacy violations, or delay release for manual review and risk operational inefficacy. To resolve this tension, we require automated systems capable of enforcing policy-aware data sharing—mechanisms that dynamically classify, evaluate, transform, and release data to ensure compliance without inducing unacceptable latency.

\subsection{Challenges of Multimodal Privacy Compliance}

Automating privacy compliance in disaster response is challenging because disaster data are diverse and context-dependent. Unlike structured database records, disaster artifacts often include unstructured media, such as images and videos, and complex metadata that capture geospatial and temporal information.
\subsubsection{Heterogeneous Data Modalities.}
 A single disaster event links textual declarations (such as incident type, location, and dates), geospatial features (such as impact area polygons), imagery, and sensor data. Each modality poses distinct privacy risks; for example, aerial imagery may contain incidental PII that is unstructured and difficult to detect~\cite{yaacoub2023survey}.
 \subsubsection{Contextual Integrity and Audience-Dependent Policies.}
 Data sensitivity often depends on the recipient. Sharing original imagery with a trusted law enforcement partner might be permissible under specific Information Sharing Access Agreements (ISAAs), whereas sharing the same imagery with the public or volunteer organizations requires strict anonymization \cite{dhs2011privacy_sharing}. Policies must be recipient-aware and activity-specific.
 \subsubsection{Dynamic Compliance States.}
 Artifacts transition between compliance states as transformations are applied. A UAS image may be non-compliant initially but becomes shareable after applying specific transforms (for example, EXIF metadata removal). The system must track the provenance of these derived artifacts to ensure obligations are met before release.
 \subsubsection{Regulatory Interplay.}
 Compliance is not governed by a single static rule but by a complex ecosystem of overlapping frameworks that span federal statutes, technical standards, and agency-specific directives. For networked sensors such as UAS, guidance such as NISTIR (National Institute of Standards and Technology Interagency Report) 8228 \cite{nistir8228} is particularly relevant; it categorizes these platforms as IoT edge devices, subjecting them to distinct cybersecurity and privacy risk management protocols.

\subsection{Our Contribution}
Prior research has demonstrated the utility of Knowledge Graphs (KGs) and Semantic Web technologies for automating regulatory compliance \cite{bonatti2020special, governatori2015legalruleml}. The \texttt{IoT-Reg} framework \cite{echenim2023iotreg} used these technologies to model privacy rules for wearable devices \cite{echenim2025kgllm}. However, existing semantic frameworks predominantly support binary access control, granting or denying access based on static attributes. This binary approach proves inadequate in disaster management scenarios, where restricting data access due to minor privacy infractions can impede prompt response efforts. In this study, we integrate deontic logic (modeling obligations, permissions, and prohibitions) directly into the operational workflow of disaster management systems, where fail-closed safety is paramount.

To address this gap, we present a Policy-Aware Decision Framework that integrates a domain-specific Disaster Management Knowledge Graph (DKG) with the regulatory intelligence of \texttt{IoT-Reg}. Our specific contributions are:

\begin{itemize}
    \item We design a comprehensive DKG that integrates real-world FEMA disaster declarations, geospatial features, and NOAA (National Oceanic and Atmospheric Administration) emergency response imagery.

    \item We extend the \texttt{IoT-Reg} ontology to map FEMA/DHS (Department of Homeland Security) disaster management concepts to privacy rules. We frame UAS and remote sensing platforms as networked Internet of Things (IoT) devices, deriving privacy risk mitigation rules from guidance such as NISTIR 8228 \cite{nistir8228} alongside General Data Protection Regulation (GDPR) and FEMA-specific mandates.

    \item We formulate a release decision function based on Deontic Logic that supports not just \textsc{Allow} or \textsc{Block} decisions, but a conditional \textsc{Allow-with-Transform} verdict. This engine bridges the gap between policy obligations and technical actions necessary to satisfy them.

    \item We implement an end-to-end system that automatically executes required transformations and logs every decision. We introduce a semantic incident detection mechanism that records blocked shares as Privacy Incidents in the KG, enabling automated auditing and reporting consistent with DHS Privacy Incident Handling Guidance \cite{dhs2024privacy_incident}. 
\end{itemize}

\noindent The remainder of this paper is organized as follows: Section 2 provides background on the disaster management domain, the \texttt{IoT-Reg} framework, relevant regulatory drivers, and the threat model and system requirements. Section 3 details the schema design of the Disaster and Policy Knowledge Graphs. Section 4 formalizes the Policy-Aware Decision Framework and its safety properties, while Section 5 describes the Operational Enforcement Layer and transformation pipelines. Section 6 presents the Analytical Monitoring Layer for compliance auditing. In Section 7, we evaluate the framework in terms of KG scale and structural validity, policy decision correctness, transform-induced state changes, and query performance. Section 8 discusses generalizability and limitations, Section 9 reviews related work, and Section 10 concludes the paper.

\section{Background and Problem Setting}

\subsection{Deontic Logic and the IoT-Reg Framework}
\label{subsec:deontic_iotreg}

Privacy regulations governing connected systems are fundamentally normative: they specify what data controllers \emph{must}, \emph{may}, and \emph{must not} do with personal data. Deontic logic provides a formal vocabulary for such norms and has been widely used to model legal and policy constraints in a machine-interpretable way~\cite{governatori2015legalruleml}.

At a high level, deontic logic introduces three primary operators over actions:
\begin{itemize}
    \item Obligation: an action ought to be performed (for example, logging disclosures of personally identifiable information (PII)).
    \item Permission: an action is allowed under specified conditions.
    \item Prohibition: an action is forbidden, often represented as an obligation not to perform that action.
\end{itemize}
These operators by themselves do not prescribe a particular enforcement strategy; they provide a semantic layer on top of which concrete decision procedures and safety conditions can be defined.

\paragraph{IoT-Reg as a Deontic Ontology for IoT Privacy.}
\texttt{IoT-Reg} is an OWL (Web Ontology Language)/Resource Description Framework (RDF) ontology that uses this deontic vocabulary to model IoT data privacy regulations in a knowledge graph form~\cite{echenim2023iotreg}. It extends standard W3C vocabularies such as the Provenance Ontology (\texttt{PROV-O})\footnote{http://www.w3.org/ns/prov\#} and the Sensor, Observation, Sample, and Actuator (\texttt{SOSA})\footnote{https://www.w3.org/ns/sosa/} ontology with concepts tailored to privacy and regulatory compliance. Conceptually, it provides:

\begin{itemize}
    \item Deontic Statement Classes: \texttt{iot-reg:Permission} for actions that are allowed under specified conditions, 
    \texttt{iot-reg:Obligation} for actions that must be carried out (such as encryption and logging), and 
    \texttt{iot-reg:Prohibition} for actions that are not allowed, modeled as subclasses of a generic Deontic Statement class.
    
    \item Lifecycle-oriented Activities: 
    Privacy and security requirements are stratified over \texttt{iot-reg:DataLifecycle} subclasses such as \texttt{DataCollection}, \texttt{DataProcessing}, \texttt{DataSharing}, \texttt{DataRetention}, and \texttt{DataDeletion}. 
    This operationalizes the specific data handling activities that NISTIR~8228 highlights as critical risk vectors, while aligning with the lifecycle models in regulations such as the GDPR.
    
    \item Data and Recipient Context:
    Deontic individuals are linked to data types (\texttt{iot-reg:PersonalData}, \texttt{iot-reg:Image}, \texttt{iot-reg:FeatureOfInterest}) and to audiences (such as internal operators and partner agencies) modeled as \texttt{iot-reg:Recipient}. 
    
    \item Regulatory Provenance:
    \texttt{IoT-Reg} associates deontic statements with their source provisions using PROV-O, linking specific permissions, prohibitions, and obligations to clauses in regulations such as GDPR and NISTIR 8228. This supports the explanation and audit of compliance decisions.
\end{itemize}

\texttt{IoT-Reg} has been used to encode NISTIR~8228 risk-mitigation goals and risk mitigation areas for wearables and to assess how vendor privacy policies align with NIST expectations for IoT devices~\cite{echenim2023wearables}. It has also been populated with GDPR scenarios and coupled with a large language model to answer both deontic (normative) and non-deontic (factual) compliance queries for IoT manufacturers~\cite{echenim2025kgllm}. In these settings, IoT-Reg serves as a reusable backbone for representing permissions, obligations, and prohibitions over IoT data-handling activities.

In this paper, \texttt{IoT-Reg} provides the policy backbone for disaster data. We reuse its deontic classes (\textsc{Permission}, \textsc{Obligation}, \textsc{Prohibition}), its lifecycle view of data activities, and its treatment of data and recipients as first-class entities. On top of this existing framework, we introduce disaster-specific policy concepts and connect them to multimodal artifacts in the DKG (Section~\ref{sec:kg_design}), and later define our system’s release decision function and safety properties as a separate contribution (Section~\ref{sec:decision_framework}).

\subsection{Regulatory Drivers}
\label{sec:regulatory_drivers}
Our system design is driven by a mesh of federal guidance that governs how disaster data must be handled. We synthesize three primary regulatory frameworks to derive the deontic logic rules for our Policy Knowledge Graph (PKG).

\subsubsection{NISTIR 8228: The IoT Privacy Baseline}
NIST Interagency Report 8228, \textit{Considerations for Managing IoT Cybersecurity and Privacy Risks} \cite{nistir8228}, frames the privacy challenge for networked sensing devices. We treat FEMA's UAS fleet as IoT edge devices: they interact with the physical world, capture data continuously, and lack traditional consent mechanisms. 

NISTIR 8228's risk mitigation goals address PII data actions, including collection, storage, processing, and transmission. Specifically, \textit{Goal 3: Protect Individuals' Privacy} highlights the need to mitigate privacy risks arising from PII processing beyond those managed by device and data security controls. Building on this foundation, we use the \texttt{IoT-Reg} framework to formalize these concepts into a stratified data lifecycle model (collection, processing, sharing, retention, and deletion), motivating our multimodal KG design, in which artifacts track their own provenance and lifecycle state independently of the device that captured them.

\subsubsection{FEMA SORN DHS/FEMA-008: Structured PII Limits}
The Disaster Recovery Assistance Files System of Records Notice (SORN) \cite{dhs2024sorn} governs the structured PII (names, addresses, financial data) collected from survivors. It establishes Routine Uses that function as conditional permissions:

\begin{itemize}
    \item Routine Use J (Unmet Needs): Permits sharing with voluntary organizations only to address specific disaster-related needs.
    \item Routine Use I (Duplication of Benefits): Permits sharing with other agencies to prevent fraud but requires strict data minimization.
\end{itemize}

\noindent We model these as audience-specific Permissions. For example, permission to share with a partner agency is conditional on encryption and logging obligations. Sharing with the public is implicitly prohibited unless a specific exception exists.

\subsubsection{FEMA UAS PIA-055: Unstructured Imagery Mandates}
The Privacy Impact Assessment (PIA) for UAS-derived imagery \cite{dhs2020uas_pia} addresses the unique risks of aerial surveillance. It acknowledges that while FEMA does not target individuals, UAS sensors inevitably capture \textit{incidental} PII.

\begin{itemize}
    \item The PIA states that any PII that is discovered in the imagery will be obfuscated or removed prior to being shared. This is a hard Obligation that must be satisfied before any public release.
    \item It forbids using UAS data to identify individuals or monitor an individual or group's movement. This serves as a global Prohibition on any processing activity involving facial recognition or tracking, overriding any other permissions.
\end{itemize}

\noindent These rules drive our \textsc{Transform}-based enforcement mechanism. A public release request triggers the required transforms to satisfy the PIA's obfuscation mandate.

\begin{figure*}[tp]
    \includegraphics[width=2\columnwidth]{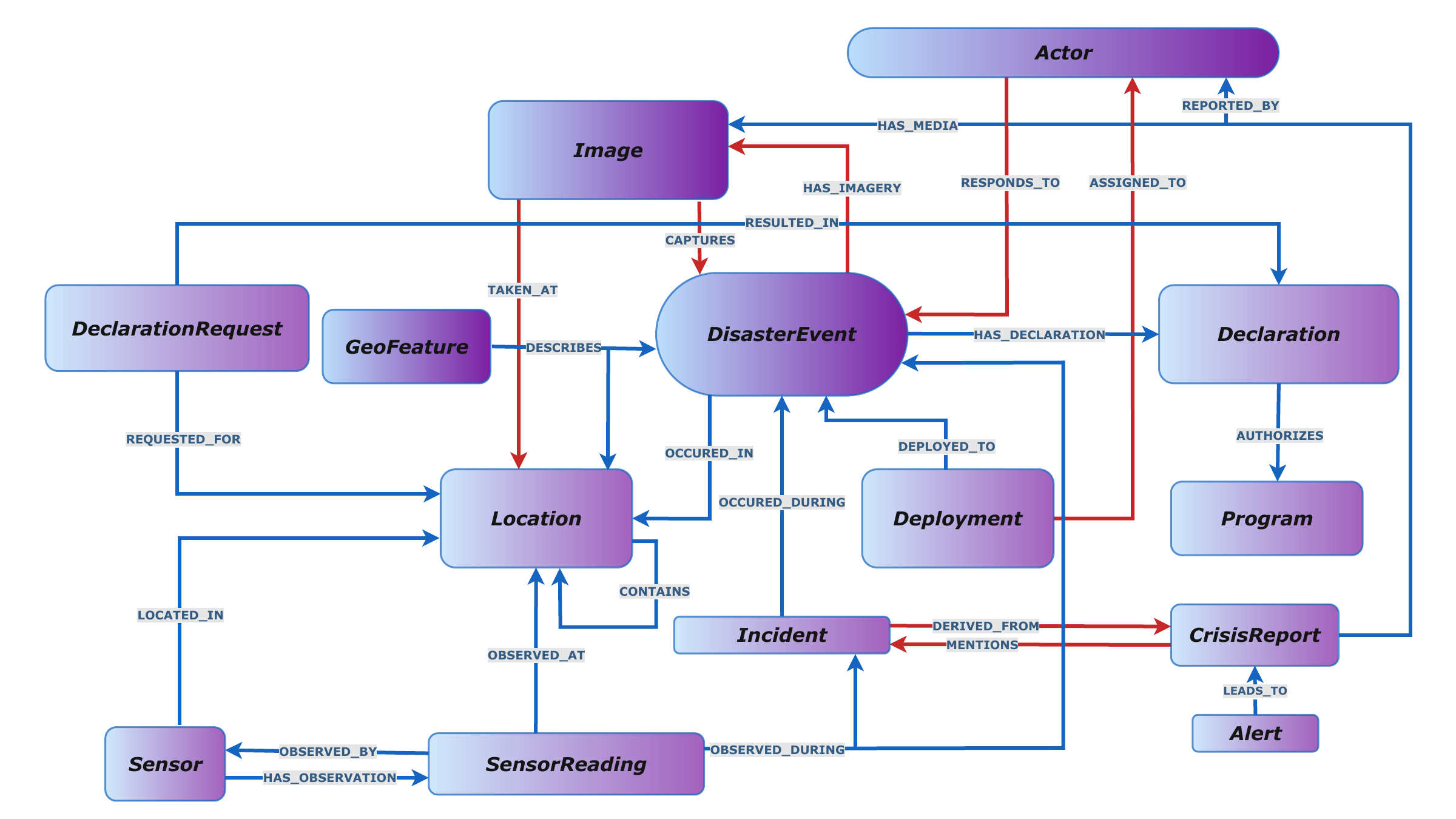}
    \caption{The Disaster Management Ontology.}
    \label{fig:dkg}
\end{figure*}

\subsection{Threat Model and System Requirements}
\label{subsec:threat_model}
\subsubsection{Threat Model}
We consider a multi-stakeholder disaster-response environment in which a central authority (e.g., FEMA) manages a DKG and disseminates artifacts to diverse recipients subject to regulatory constraints. Following applied security methodology, we ground our threat enumeration in a model of the adversary's resources and intent. \cite{Do2019AdversaryModel}. Our threat model covers adversarial behavior and operational failures.

\paragraph{Trusted Computing Base (TCB)} We assume the integrity of the KG infrastructure and access controls, the ontology and PKG definitions, the policy engine logic, and trusted transformation services. Write access to compliance-critical assertions is mediated strictly by trusted services; direct updates to the triple store by end users are disallowed.

\paragraph{Adversary Model.} We consider (i) honest-but-curious authorized partners that follow protocol but attempt to infer sensitive information from permitted releases, and (ii) malicious insiders (or compromised accounts) that attempt to bypass compliance controls to exfiltrate PII.

\paragraph{Threat taxonomy.} We structure privacy threats using LINDDUN \cite{deng2011privacy} and integrity threats using STRIDE \cite{shostack2014threat}. We specifically address the following adversary capabilities:

\begin{itemize}
    \item[A1] \textit{Over-scoped sharing} (STRIDE: Elevation; LINDDUN: Non-compliance). Authorized users may attempt to share data beyond their permitted scope, either maliciously or due to policy confusion (e.g., releasing original UAS imagery without anonymization).
    \item[A2] \textit{Semantic ambiguity exploitation} (LINDDUN: Unawareness /Non-compliance). Adversaries may exploit ambiguity in natural-language policies to justify unauthorized access, for example, by interpreting partner agencies broadly to include unauthorized contractors.
    \item[A3] \textit{Sanitization/obligation bypass} (LINDDUN: Identifiability/Disclosure). Actors may attempt to release artifacts without the required transformations or claim that an artifact is anonymized (for instance) when it still contains sensitive visual PII.
    \item[A4] \textit{Metadata forgery} (STRIDE: Tampering). Adversaries may attempt to inject false metadata on artifacts to bypass automated compliance checks.
    \item[A5] \textit{Action repudiation} (STRIDE: Repudiation). An insider may dispute having initiated a non-compliant release. Without tamper-evident audit records, accountability fails.
\end{itemize}

\noindent \textit{Operational Failures.} In addition to adversarial threats, the system must address non-malicious failures that impact compliance:

\begin{itemize}
    \item[F1] \textit{Policy drift.} As regulations and guidance change, static hard-coded rules may become obsolete, leading to noncompliant shares.
    \item[F2] \textit{Silent transform failure.} Automated transforms may fail silently, resulting in artifacts that appear compliant but still contain PII.
    \item[F3] \textit{Evaluation/query failures.} Query outages or policy-evaluation failures could otherwise lead to fail-open behavior if not handled explicitly.
\end{itemize}

\noindent We exclude network-layer denial-of-service, physical compromise of the data center, and steganographic exfiltration.

\subsubsection{System Requirements}
To mitigate these threats and align with regulatory drivers (defined in Section \ref{sec:regulatory_drivers}), the system must satisfy the following requirements:

\begin{itemize}
    \item[R1] \textit{Prohibition Dominance} (defense against A1, A2). Explicit prohibitions must override permissions. If a prohibition matches a request, the verdict must be \textsc{Block} regardless of other permissions.

    \item[R2] \textit{Fail-Closed Default} (defense against F3). The system must enforce a deny-by-default posture. If policy queries fail, the PKG is unreachable, or no explicit permission matches the request tuple, the verdict must default to \textsc{Block}.

    \item[R3] \textit{Verifiable Obligation Satisfaction} (defense against A3, F2). Obligations must be verified against the artifact state recorded in the DKG, not user assertions. The system must map obligations to verifiable compliance flags and require that they be met before release.

    \item[R4] \textit{Provenance Binding and Tamper-Evident Audit} (defense against A4, A5). Derived artifacts must link to originals via \texttt{prov:}~\cite{w3c_provo}. Compliance assertions must be produced only by trusted transformation pipelines (not manual user input), and release decisions must be recorded in tamper-evident audit logs to support accountability.

    \item[R5] \textit{Semantic Incident Detection} (defense against A1, F1). Blocked or anomalous requests must be logged as privacy incidents in the DKG to support audit and response workflows consistent with DHS guidance \cite{dhs2024privacy_incident}. Incident records should classify the reason (e.g., \textsc{Prohibited\_Share}, \textsc{No\_Permission}) to support policy refinement.

    \item[R6] \textit{Audience Binding} (defense against A2). Decisions must bind each request to an authenticated audience class (e.g., Public vs. Partner) and enforce recipient-dependent rules to prevent context collapse across release channels.
\end{itemize}

\section{Knowledge Graph Design}
\label{sec:kg_design}

We realize policy-aware disaster data sharing through two interacting RDF graphs. The Disaster Management Knowledge Graph (DKG) models multimodal disaster data, while the Policy Knowledge Graph (PKG) encodes FEMA/DHS privacy and data sharing rules using \texttt{IoT-Reg} and an extension namespace. The release decision function (Section~\ref{sec:decision_framework}) queries both graphs: the DKG exposes the current state of artifacts, and the PKG supplies the applicable deontic logic.

\subsection{Disaster Management KG Schema}
\label{sec:dkg-schema}

The DKG is defined in the \texttt{disaster\_mgt} (\texttt{dm:}) namespace and is anchored in an authoritative, text-centric layer derived mostly from FEMA disaster declaration datasets. This textual layer serves as a backbone to which imagery, geospatial features, and future modalities attach deterministically via identifiers, locations, and time spans.
The main textual classes are \texttt{dm:DisasterEvent}, \texttt{dm:Declaration}, \texttt{dm:DeclarationRequest}, \texttt{dm:Program}, and \texttt{dm:Location}.

\begin{itemize}
    \item \texttt{dm:DisasterEvent} represents the principal crisis (e.g., a named hurricane, wildfire, flood) and key attributes include identifiers (\texttt{dm:disasterNumber}), hazard descriptors (\texttt{dm:incidentType}, \texttt{dm:declarationTitle}), and temporal anchors (\texttt{dm:incidentBeginDate}, \texttt{dm:incidentEndDate}, \texttt{dm:declarationDate}). Each event links to at least one \texttt{dm:Location} via \texttt{dm:occured\_in} and, when declared, to a \texttt{dm:Declaration} via \texttt{dm:has\_declaration}.
    
    \item \texttt{dm:Declaration} captures official government responses. Attributes such as \texttt{dm:femaDeclarationString}, \texttt{dm:declaration Type}, \texttt{dm:declarationDate}, and \texttt{dm:disasterCloseoutDate} model the declaration lifecycle, while \texttt{dm:authorizes} links declarations to \texttt{dm:Program} individuals that represent Individual Assistance (IA), Public Assistance (PA), Hazard Mitigation (HM), and related programs~\cite{openfema_declarations}.
    
    \item \texttt{dm:DeclarationRequest} represents formal requests that may or may not result in declarations. Attributes such as \texttt{dm:declarationRequestNumber}, requested incident interval, requested assistance flags, and \texttt{dm:requestStatus} allow the graph to retain latent, unmet demand even when no federal declaration is issued~\cite{openfema_denials}.
    \item \texttt{dm:Program} models forms of assistance, enabling queries about which programs were available for a given event or location.
    
    \item \texttt{dm:Location} encodes geographic entities (states, counties, tribal areas) with attributes such as \texttt{dm:placeName}, \texttt{dm:stateName}, and \texttt{dm:placeCode}, and participates in a hierarchy via \texttt{dm:contains}.
\end{itemize}

Figure~\ref{fig:dkg} shows our Disaster Management Ontology.
These classes are populated from four FEMA open datasets: \textit{Disaster Declarations Summaries}, \textit{Web Declaration Areas}, \textit{Declaration Denials}, and \textit{Web Disaster Summaries}~\cite{openfema_declarations,openfema_web_areas,openfema_denials,openfema_web_summaries}. We follow a mapping strategy that ingests a column only if it contributes to identity, location, time, or assistance modality.

The schema also introduces \texttt{dm:CrisisReport} (with subclasses such as \texttt{dm:FieldReport} and \texttt{dm:SocialMediaPost}), \texttt{dm:Alert}, \texttt{dm:Actor}, \texttt{dm:Sensor}, and \texttt{dm:SensorReading}. These classes support future ingestion of additional data sources such as IPAWS alerts \cite{openfema_ipaws}, streaming sensor feeds, crowdsourced reports, and other public datasets but are not populated in the current prototype.

\paragraph{Geospatial features and Imagery.}
To support multimodal reasoning, the textual backbone is extended with \texttt{dm:GeoFeature} and \texttt{dm:Image} as primary classes.

\begin{itemize}
    \item \texttt{dm:GeoFeature} represents geographic footprints of impact areas, populated from web declaration feeds and shapefile endpoints. Attributes include \texttt{geo:asWKT} for geometry (normalized to WGS84\footnote{https://www.w3.org/2003/01/geo/} and encoded using the Open Geospatial Consortium (OGC) GeoSPARQL standard \cite{ogc_geosparql}) and derived summaries such as \texttt{dm:hasAreaSqKm}. Each \texttt{dm:GeoFeature} links to the event or location it describes via \texttt{dm:describes}.
    \item \texttt{dm:Image} represents UAS and other emergency response imagery. Current instances are derived from NOAA Emergency Response Imagery metadata \cite{noaa_eri} and are dual-typed as \texttt{dm:Image} and \texttt{iot-reg:Image}. Attributes include \texttt{dm:fileUrl}, \texttt{dm:captured\_time}, and optional geospatial keys; images are linked to events and locations via \texttt{dm:captures} and \texttt{dm:taken\_at}, respectively.
\end{itemize}

\paragraph{Privacy flags.}
Images carry boolean flags that encode compliance-relevant state, which are schema-attached rather than free-form annotations:
\begin{itemize}
    \item \texttt{dm:containsPII}: Indicates whether the artifact is believed to contain PII.
    \item \texttt{iot-reg:isAnonymized}: Indicates whether PII-obfuscation transforms have been applied.
    \item \texttt{iot-reg:isEncrypted}: Indicates whether the stored artifact is encrypted.
    \item \texttt{dm:isRetained}: Indicates whether the artifact is held as a long-term record.
\end{itemize}
These flags are designed to be set programmatically by transformation services in the operational layer, not directly edited by users.

\subsection{Policy KG: FEMA/DHS Rules in IoT-Reg}
\label{sec:policy-kg}

\begin{figure}
    \centering
    \includegraphics[scale=0.5
    ]{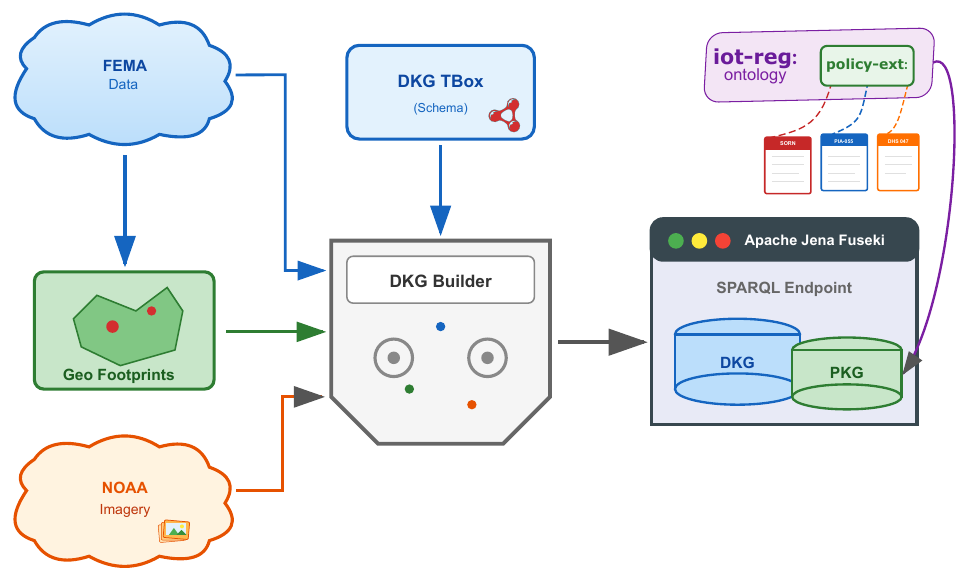} 
    \caption{Knowledge graph population pipeline showing external data sources, the DKG Builder processing component, and the resulting SPARQL endpoint hosting both the Disaster KG and Policy KG, with policy rules derived from FEMA/DHS regulatory documents.}
    \label{fig:kp_pop_ingestion}
\end{figure}

The Policy KG reuses the \texttt{IoT-Reg} ontology~\cite{echenim2023iotreg} to model FEMA and DHS privacy requirements as deontic statements, and extends it with a \texttt{policy-ext} namespace. It is maintained as a separate RDF graph and populated manually from the FEMA SORN DHS/FEMA-008~\cite{dhs2024sorn}, FEMA UAS PIA-055~\cite{dhs2020uas_pia}, and DHS Privacy Incident Handling Instructions~\cite{dhs2024privacy_incident}. Figure~\ref{fig:kp_pop_ingestion} illustrates this population pipeline.

\paragraph{Reuse of IoT-Reg.}
IoT-Reg provides the base classes: \texttt{iot-reg:Regulation}, \texttt{iot-reg:Permission}, \texttt{iot-reg: Prohibition}, \texttt{iot-reg:Obligation}, and \texttt{iot-reg:Recipient}. In our PKG, a central individual \texttt{policy-ext:FEMA\_Controller} (an instance of \texttt{iot-reg:Controller}) represents FEMA as the data controller, and \texttt{policy-ext:FEMA\_DisasterDataPolicy} aggregates the relevant rules via \texttt{iot-reg:hasPermission}, \texttt{iot-reg:hasProhibition}, and \texttt{iot-reg:hasObligation}.

\paragraph{policy-ext extensions.}
The deontic rules were derived by systematically encoding each permission, prohibition, and obligation statement from the three source documents that constrain artifact sharing by audience or data type. To connect the rules to artifact states in the DKG, we define the following properties:
\begin{itemize}
    \item \texttt{policy-ext:concernsData}: Refines its parent property, \texttt{iot-reg:involvesPersonalData}, to address both traditional PII records and imagery.
    \item \texttt{policy-ext:requiresTransform}: A datatype property on \texttt{iot-reg:Obligation} listing transform function names (e.g., \texttt{strip\_exif}) that satisfy the obligation.
    \item \texttt{policy-ext:checksFlag}: A datatype property naming the boolean flag in the DKG (e.g., \texttt{iot-reg:isAnonymized}) that witnesses satisfaction.
\end{itemize}

\subsection{Provenance and Incident Modeling}
\label{sec:provenance-pii}

To make compliance states auditable, we treat provenance and incidents as first-class graph entities rather than opaque log entries.

\paragraph{Artifact derivation and state.} 
The DKG follows the W3C PROV-O standard~\cite{w3c_provo} to represent transformations. When transforms are applied, we represent them by creating a new \texttt{dm:Image} instance instead of mutating the original.
The derived image is linked to the original via \texttt{prov:wasDerivedFrom}, records a creation time via \texttt{prov:generatedAtTime}, and carries updated privacy flags.

\paragraph{Incidents and audits.}
Privacy incidents are represented as \texttt{iot-reg:PersonalDataBreach} individuals in the DKG. Each incident carries annotations such as \texttt{dm:incidentCategory} and \texttt{dm:incidentReason}. Audit activities are modeled as \texttt{policy-ext:IncidentAudit} instances, connected to incidents via \texttt{dm:auditsIncident}.

\section{Policy-Aware Decision Framework}
\label{sec:decision_framework}
In this section, we formalize the problem of automated privacy compliance in disaster response as a release decision function. We define the interaction between the Disaster Management Knowledge Graph (DKG) and the Policy Knowledge Graph (PKG) and establish the safety properties necessary to ensure fail-safe operations.

\subsection{Problem Formation}
We model the compliance verification process as a deterministic function over two dynamic knowledge graphs. Let $\mathcal{D}$ denote the DKG, which stores multimodal artifacts, their metadata, and provenance. Let $\mathcal{P}$ denote the PKG, which encodes regulatory rules as deontic logic constructs (permissions, obligations, prohibitions).

We define a release request as a tuple $r = \langle a, u, \alpha, \tau \rangle$, where $a \in \mathcal{D}$ is the unique identifier (URI) of the artifact requested, $u \in \mathcal{P}_{\text{Actors}}$ is the target recipient or audience (e.g., \texttt{policy-ext:PublicAudience}), $\alpha \in \mathcal{P}_{\text{Activities}}$ is the intended data lifecycle activity (e.g., \texttt{iot-reg:DataSharing}), and $\tau$ is the abstract data type of $a$ (e.g., \texttt{iot-reg:PersonalData}), derived via type inference on $\mathcal{D}$.

The release decision function $\delta$ maps the current state of the
graphs and the request to a compliance verdict and a set of required transformations:
$$
\delta(\mathcal{D}, \mathcal{P}, r) \to \langle v, T \rangle
$$
where $v \in \{\textsc{Allow}, \textsc{Block}, \textsc{Allow-with-Transform}\}$ is the decision verdict, and $T$ is the ordered set of transformation functions required to satisfy pending obligations. When $v = \textsc{Allow}$, $T = \emptyset$ and the original artifact $a$ may be released. When $v = \textsc{Allow-with-Transform}$, the enforcement layer (Section~\ref{sec:enforcement}) applies $T$ to produce a derived artifact $a'$, which is linked to $a$ via provenance and verified for compliance before release.

\subsection{Compliance Logic and Safety Properties}
To ensure the system adheres to regulatory standards, the decision function $\delta$ must satisfy three axiomatic safety properties.

\subsubsection{Property 1: Prohibition Dominance.}
Explicit prohibitions override all permissions. If the policy graph $\mathcal{P}$ contains a prohibition rule $\rho$ that matches the tuple $\langle u, \alpha, \tau \rangle$, the verdict must be \textsc{Block}, regardless of any existing permissions.

\subsubsection{Property 2: Fail-Closed Default.}
Access is denied by default. A verdict of \textsc{Allow} or \textsc{Allow-with-Transform} is generated only if an explicit permission $\pi \in \mathcal{P}$ exists that matches $r$. If no such permission is found, or if query execution against $\mathcal{D}$ or $\mathcal{P}$ fails, the system defaults to $v = \textsc{Block}$.

\subsubsection{Property 3: Obligation Consistency.}
A permission $\pi$ is valid only if its attached obligations are satisfied. For an obligation $o$ linked to permission $\pi$, let $f_o$ be the compliance flag property in $\mathcal{D}$ tracked by $o$.

\begin{itemize}
    \item If $a$ already satisfies $f_o$ (i.e., $\mathcal{D} \models f_o(a)$), the obligation is met.
    \item If $a$ does not satisfy $f_o$, the system must identify a transformation $t \in \mathcal{T}$ capable of asserting $f_o$.
    \item If no such transformation exists in $\mathcal{P}$, the verdict falls back to \textsc{Block}.
\end{itemize}

\subsection{Algorithm Design}
We define the compliance framework by two primary algorithms: the decision logic and the incident logging mechanism.

\subsubsection{Algorithm~\ref{alg:policy_aware}: Policy-Aware Release Decision}
Algorithm 1 details the \texttt{Release Decision} procedure. It prioritizes prohibition checks to ensure fail-fast behavior. It subsequently attempts to match permissions and resolves unsatisfied obligations by querying the Policy KG for mapped transformation functions.

\begin{algorithm}
\caption{Policy-Aware Release Decision}
\label{alg:policy_aware}
\begin{algorithmic}[1]

\Require Disaster KG $\mathcal{D}$, Policy KG $\mathcal{P}$, Request $r = \langle a, u, \alpha, \tau \rangle$
\Ensure Verdict $v$, Transform Set $\mathcal{T}$

\Statex 
\textbf{Definitions:}
\Statex $\Phi(p)$: Retrieves obligations attached to permission $p$
\Statex $\text{Flag}(o)$: Retrieves the boolean property checked by obligation $o$
\Statex $\text{Func}(o)$: Retrieves the transform function satisfying $o$
\Statex

\State \textbf{Phase 1: Check Prohibitions (Fail-Fast)}
\State $\mathcal{R}_{prohib} \gets \text{Query}(\mathcal{P}, \text{Prohibitions matching } u, \alpha, \tau)$
\If{$\mathcal{R}_{prohib} \neq \emptyset$}
    \State \Return $\langle \textsc{Block}, \emptyset \rangle$
\EndIf

\Statex
\State \textbf{Phase 2: Check Permissions (Fail-Closed)}
\State $\mathcal{R}_{perm} \gets \text{Query}(\mathcal{P}, \text{Permissions matching } u, \alpha, \tau)$
\If{$\mathcal{R}_{perm} = \emptyset$}
    \State \Return $\langle \textsc{Block}, \emptyset \rangle$
\EndIf

\Statex
\State \textbf{Phase 3: Evaluate Obligations}
\State Select highest priority permission $p \in \mathcal{R}_{perm}$
\State $\mathcal{T} \gets \emptyset$

\For{\textbf{each} obligation $o \in \Phi(p)$}
    \State $flag \gets \text{Flag}(o)$
    \If{$\text{Query}(\mathcal{D}, \neg \text{HasFlag}(a, flag))$} \Comment{Check if asset violates obligation}
        \If{$\exists \text{Func}(o)$}
            \State $\mathcal{T} \gets \mathcal{T} \cup \{\text{Func}(o)\}$ \Comment{Map remediation function}
        \Else
            \State \Return $\langle \textsc{Block}, \emptyset \rangle$ \Comment{Obligation unsatisfiable}
        \EndIf
    \EndIf
\EndFor

\Statex
\State \textbf{Phase 4: Final Verdict}
\If{$\mathcal{T} = \emptyset$}
    \State \Return $\langle \textsc{Allow}, \emptyset \rangle$
\Else
    \State \Return $\langle \textsc{Allow-with-Transform}, \mathcal{T} \rangle$
\EndIf

\end{algorithmic}
\end{algorithm}

\subsubsection{Algorithm~\ref{alg:incident_registration}: Incident Registration}
To satisfy the regulatory accountability requirements (of GDPR and NISTIR 8228), blocked requests involving personal data must be immutably recorded. Algorithm 2 defines the \texttt{Log Incident} procedure, which inserts a \texttt{PersonalDataBreach} instance into the DKG if a request is blocked due to policy violations.

\begin{algorithm}
\caption{Incident Registration}
\label{alg:incident_registration}
\begin{algorithmic}[1]

\Require Disaster KG $\mathcal{D}$, Request $r$, Verdict $v$, Reason $msg$
\Ensure Incident URI $i$

\Statex
\State \textbf{Phase 1: Condition Check}
\If{$v \neq \textsc{Block} \lor r.\tau \neq \texttt{iot-reg:PersonalData}$}
    \State \Return $\emptyset$ \Comment{No privacy incident to log}
\EndIf

\Statex
\State \textbf{Phase 2: Classify Incident Category}
\State $c \gets \textsc{Classify}(msg)$
\Statex \Comment{Maps reason to $\{\textsc{Prohibited\_Share}, \textsc{No\_Permission}, \textsc{Invalid\_Audience}, \newline \textsc{Transform\_Failure}, \textsc{Other}\}$}

\Statex
\State \textbf{Phase 3: Construct Knowledge Graph Triplets}
\State $i \gets \text{NewURI}(\mathcal{D}, \textsc{Incident})$
\State $t_{\text{now}} \gets \text{CurrentTimestamp}()$

\State $\text{Insert}(\mathcal{D}, \langle i, \text{rdf:type}, \text{iot-reg:PersonalDataBreach} \rangle)$
\State $\text{Insert}(\mathcal{D}, \langle i, \text{prov:wasDerivedFrom}, r.a \rangle)$
\State $\text{Insert}(\mathcal{D}, \langle i, \text{dm:incidentCategory}, c \rangle)$
\State $\text{Insert}(\mathcal{D}, \langle i, \text{dm:incidentReason}, msg \rangle)$
\State $\text{Insert}(\mathcal{D}, \langle i, \text{dm:incidentDetectedAt}, t_{\text{now}} \rangle)$

\State \Return $i$

\end{algorithmic}
\end{algorithm}

\section{Operational Enforcement Layer}
\label{sec:enforcement}

\begin{figure*}[tp]
    \centering
    \includegraphics[scale=0.5
    ]{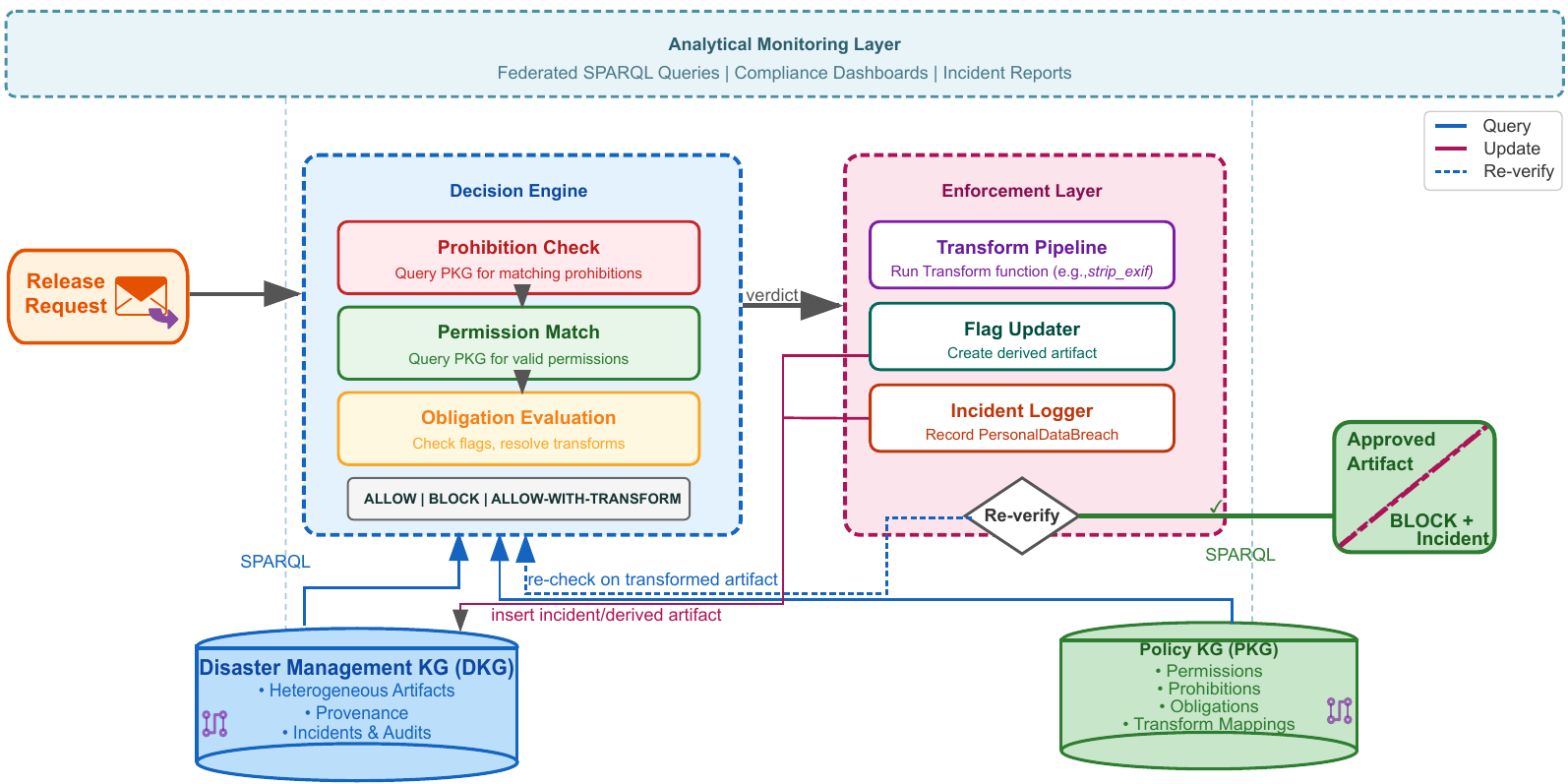} 
    \caption{End-to-end architecture of the privacy-aware disaster data sharing framework, showing the flow from release requests through policy evaluation, transform execution, and compliance verification, with the Disaster KG and Policy KG as dual knowledge sources.}
    \label{fig:system_architecture}
\end{figure*}

 We implement the policy-aware decision framework as an operational enforcement layer that takes structured release requests, invokes the decision algorithms from Section~\ref{sec:decision_framework}, executes required transforms, and updates the Disaster KG with derived artifacts and incident records. See Figure~\ref{fig:system_architecture}. This section describes the packet abstraction, the transform pipeline, the artifact derivation workflow, and the incident logging mechanism.

\subsection{Packet Model for Release Requests}
\label{sec:packet-model}

Release decisions are driven by request packets that encode the context of a proposed share without exposing original files to the policy engine. Each packet is a JSON document processed by the \texttt{Release Decision Runner}:

\begin{itemize}
    \item \textit{Artifact identifier:} A URI $a$ that must already exist in the Disaster KG (e.g., a source UAS image or a derived encrypted artifact).
    \item \textit{Audience and activity:} The intended recipient $u$ (e.g., \texttt{policy-ext:PublicAudience}) and the lifecycle activity $\alpha$ (focusing on \texttt{iot-reg:DataSharing}).
    \item \textit{Data type:} An asserted or inferred data type $\tau$. When the packet omits $\tau$, the \texttt{DataType Detector} queries the Disaster KG to infer it from RDF types.
    \item \textit{File location:} An optional file URL or local path that the enforcement layer uses to fetch the artifact for transformation.
\end{itemize}

The packet, therefore, acts as the running counterpart of the abstract request $r = \langle a, u, \alpha, \tau \rangle$ used in the decision logic. The runner resolves missing fields (e.g., infers $\tau$), then calls the decision function through the \texttt{Compliance Policy Adapter}, which issues SPARQL queries over the PKG and DKG. If these queries fail, or if no matching permission exists, the verdict remains \textsc{Block} in line with the fail-closed property. Audience authentication is handled externally; the enforcement layer consumes authenticated audience identifiers and does not itself implement identity management.

\subsection{Transform Pipeline}
\label{sec:transform-pipeline}

When Algorithm~\ref{alg:policy_aware} returns a verdict of \textsc{Allow-with-Transform} along with an ordered set of required transforms $T$, the enforcement layer invokes a transform pipeline over the underlying file. Each transform is a deterministic function that takes a local path and returns a new path and a success flag; transforms operate on local copies resolved by a \texttt{File Resolver}.

In our prototype, transforms are modular functions invoked by the enforcement layer. We define two families:\\
\textit{Anonymization:} For UAS imagery intended for public release, the prototype implements \texttt{strip\_exif}, which removes EXIF metadata using standard image libraries. This function satisfies the \texttt{policy-ext:requiresTransform} property associated with \texttt{Oblig\_ObfuscatePII}. Deployments that require visual PII obfuscation (e.g., face blurring) can integrate additional computer-vision transforms without modifying the decision logic.\\
\textit{Encryption:} For PII shares to partner agencies, the prototype implements \texttt{encrypt\_file} using Fernet symmetric encryption (AES-128-CBC with HMAC). The encrypted artifact is packaged with metadata; the decryption key is stored separately. This satisfies the encryption obligation required by the FEMA SORN~\cite{dhs2024sorn} and DHS incident handling guidance~\cite{dhs2024privacy_incident}.

Our decision framework determines when transforms are required and verifies post-transform compliance—not the transforms themselves, which are domain-specific and pluggable. The transforms are applied in the order returned by the decision logic. If any required transform fails (e.g., due to missing libraries or corrupted files), the runner treats the request as non-compliant, returns a \textsc{Block} verdict, and records the failure in the decision log. No partial or best-effort release is attempted, preserving the fail-closed behavior specified in Section~\ref{sec:decision_framework}.

\subsection{Artifact Derivation and Flag Updates}
\label{sec:artifact-derivation}

To maintain immutable provenance (Requirement R4), the enforcement layer never mutates artifacts in the Disaster KG. Instead, when required transforms succeed, the \texttt{FlagUpdater} creates a new RDF individual that represents the transformed artifact and links it to the original.

For anonymization, the system constructs a derived URI $a'$ (e.g., appending an \texttt{\_anonymized} suffix), types it as \texttt{iot-reg:Image} and \texttt{iot-reg:FeatureOfInterest}, and updates privacy flags to reflect its state. Here, \texttt{dm:containsPII} is set to \texttt{false}, \texttt{dm:isRetained} set to \texttt{true}, \texttt{iot-reg:isAnonymized} set to \texttt{true}, and \texttt{iot-reg:isEncrypted} set to \texttt{false}.

The derived artifact is linked back to the original via \texttt{prov:wasDerivedFrom}, and annotated with \texttt{prov:generatedAtTime}, \texttt{dm:transformedBy} (a universally unique decision identifier), \texttt{dm:appliedTransforms}, and an updated \texttt{dm:fileUrl}.

For encrypted partner shares, the derived artifact is typed as both \texttt{iot-reg:Image} and \texttt{iot-reg:PersonalData}, since PII remains present. In this case, \texttt{dm:containsPII} stays \texttt{true}, \texttt{iot-reg:isEncrypted} is set to \texttt{true}, and an additional \texttt{iot-reg:usesEncryptionMethod} literal documents the cryptographic scheme. As with anonymized derivatives, the provenance chain and transform metadata are inserted into the Disaster KG.

\paragraph{Verification.} After inserting $a'$, the runner \textit{re-invokes} the decision function on the derived artifact with the same audience and activity. It verifies that the new flags in the KG satisfy all obligations and that no prohibitions apply. Only when the re-check returns \textsc{Allow} does the runner return $a'$ as the approved artifact. This validates Property 3 (Obligation Consistency): every allowed share corresponds to an artifact whose compliance state is explicitly represented in the KG and independently verified.

\subsection{Incident Detection and Logging}
\label{sec:incident-logging}

Algorithm~\ref{alg:incident_registration} specifies when a blocked request must be recorded as a privacy incident. The operational layer implements this logic via an \texttt{Incident Logger} that interacts with the Disaster KG.

When a decision produces a \textsc{Block} verdict for an artifact whose type includes \texttt{iot-reg:PersonalData}, the logger classifies the reason (e.g., prohibited share, no matching permission, transform failure) and inserts an incident instance into the KG. Each incident is typed as \texttt{iot-reg:PersonalDataBreach}, linked to the affected artifact via \texttt{prov:wasDerivedFrom}, and annotated with \texttt{dm:incidentCategory} for structured classification, \texttt{dm:incidentReason} carrying the textual explanation from the decision engine,and \texttt{dm:incidentDetectedAt} for the detection timestamp.

Follow-up audits are represented as \texttt{policy-ext:IncidentAudit} individuals linked to incidents via \texttt{dm:auditsIncident}, with \texttt{dm:auditTime} and \texttt{dm:auditNotes} fields. Because incidents and audits share the same KG as disaster artifacts and policy rules, the analytical layer (Section~\ref{sec:monitoring-layer}) can issue SPARQL queries that summarize incident categories, identify unresolved incidents, and trace how prohibited or failed shares were handled. This ties the abstract incident registration logic from Section~\ref{sec:decision_framework} to material, queryable evidence of compliance.

\subsection{Implementation Stack}
\label{subsec:implementation_stack}
The operational enforcement layer is implemented as a Python middleware service that orchestrates policy evaluation and transformations. The DKG and PKG are deployed as separate Apache Jena Fuseki datasets backed by its Triple Database (TDB2), providing transactional SPARQL 1.1 query and update over persistent storage \cite{apache_jena}. Evaluation and enforcement components interact with both graphs via SPARQL endpoints, decoupling policy logic from data persistence.

\section{Analytical Monitoring Layer}
\label{sec:monitoring-layer}

While the operational enforcement layer handles per-request release decisions, analysts require a global view of facts, compliance states, pending obligations, and privacy incidents. We provide this through an analytical monitoring layer built on federated SPARQL queries over the DKG and PKG. This layer does not re-implement decision logic; instead, it reads the same deontic rules and artifact flags to produce live compliance dashboards.

\subsection{Federated Policy--Data Queries}
\label{sec:federated-queries}

The DKG and PKG are deployed as separate SPARQL endpoints: the DKG stores operational data (disasters, imagery, privacy flags, incidents, geofeatures, and audits), and the PKG stores deontic rules (permissions, prohibitions, obligations). Analytical queries use \texttt{SERVICE} clauses to join both graphs into a single view.

Each monitoring query follows a common pattern:
\begin{itemize}
    \item Query the DKG for artifacts, privacy flags, and provenance.
    \item Query the PKG for relevant permissions and obligations, retrieving their \texttt{policy-ext:checksFlag} properties.
    \item Join both subqueries on data type and audience to compute a compliance summary.
\end{itemize}

Listing~\ref{lst:federated_query} illustrates this pattern for the global compliance dashboard. This design maintains the PKG as the single source of truth for rules and the DKG for data state, simplifying maintenance when regulations evolve.

\begin{lstlisting}[language=SPARQL, caption={Federated SPARQL query determining compliance status.}, label={lst:federated_query}]
SELECT ?imgName ?audience ?status WHERE {
  SERVICE <http://dkg-endpoint/sparql> {
    ?img a iot-reg:Image ; 
         iot-reg:isAnonymized ?isAnon .
    BIND(strafter(str(?img), '#') AS ?imgName)
  }
  SERVICE <http://pkg-endpoint/sparql> {
    ?perm iot-reg:hasRecipient ?recipient ;
          iot-reg:hasObligation ?oblig .
    ?oblig policy-ext:checksFlag ?flagProp .
    BIND(strafter(str(?recipient), '#') AS ?audience)
  }
  BIND(
    IF(?flagProp = "iot-reg:isAnonymized",
       IF(BOUND(?isAnon) && ?isAnon, "COMPLIANT", "NON_COMPLIANT"),
       "UNKNOWN"
    ) AS ?status
  )
}
\end{lstlisting}

\subsection{Compliance Dashboards and Audience Views}
\label{sec:dashboards}

We define a library of federated query templates that power dashboard views: (i) global compliance status, listing each artifact's compliance state per audience; (ii) images needing transforms, identifying non-compliant images and their required remediation; (iii) audience-specific shareability, answering "which images can be shared with audience $u$ now?"; (iv) cross-audience compliance matrix, showing where different artifact versions are needed; and (v) policy-based explanation, returning the specific prohibition or unsatisfied obligation blocking a request. Each template dynamically joins PKG rules with DKG state, enabling filterable, real-time compliance views.

\subsection{Incident and Audit Analytics}
\label{sec:incident-analytics}

The same federated approach underpins incident analytics. Templates identify unaudited \texttt{iot-reg:PersonalDataBreach} instances, retrieve audit histories for specific incidents, and aggregate incidents by \texttt{dm:incidentCategory} to surface systematic issues. Explanation queries combine DKG flags with PKG rules to trace why a specific request was blocked. \\

\noindent These templates turn the KGs into a live monitoring surface for oversight and policy tuning; performance is evaluated in Section~\ref{sec:evaluation}.

\section{Evaluation}
\label{sec:evaluation}

We evaluate the framework across four axes: the scale and structural quality of the DKG, the correctness of release decisions relative to the specified policy, the impact of transforms on compliance states within the DKG, and the query performance for single-KG and federated templates.

\subsection{Experimental Setup and Datasets}
\label{subsec:experimental_setup}
We evaluate on a fixed snapshot of the DKG and PKG spanning FEMA-recorded disasters from 2005--2024. The DKG aggregates the four OpenFEMA datasets and NOAA imagery sources described in Section~\ref{sec:kg_design}. All experiments were conducted on the stack in Section~\ref{subsec:implementation_stack}, with the PKG instantiated using the \texttt{IoT-Reg} ontology and 15 domain-specific deontic statements derived from the FEMA/DHS privacy documents.

\begin{table}[t]
\caption{Scale of the Disaster Management and Policy Knowledge Graphs, 
including class instance counts and structural quality assurance checks 
verifying referential integrity and flag consistency across the evaluation snapshot.}
\label{tab:kg-scale-qa}
\centering
\small
\begin{tabular}{@{}lr|lr@{}}
\toprule
\multicolumn{2}{c|}{\textbf{Scale Metrics}} & \multicolumn{2}{c}{\textbf{QA Checks}} \\
\midrule
DisasterEvent & 5,060 & Events w/o locations & 0 \\
Declaration & 5,060 & Decl.\ w/o events & 0 \\
DeclarationRequest & 1,257 & Derived img w/o prov & 0 \\
Location & 5,935 & Conflicting flags & 0 \\
GeoFeature & 2,690 & & \\
Image & 316,090 & & \\
\midrule
\textbf{DKG triples} & \textbf{5,102,618} & & \\
\midrule
Deontic individuals & 15 & & \\
Schema classes/props & 47 & & \\
\bottomrule
\end{tabular}
\end{table}

Table~\ref{tab:kg-scale-qa} summarizes the size of the KGs used in evaluation. The DKG snapshot contains 5{,}060 \texttt{dm:DisasterEvent} individuals, 5{,}060 \texttt{dm:Declaration} individuals, 1{,}257 \texttt{dm:DeclarationRequest} individuals, 5{,}935 \texttt{dm:Location} individuals, 2{,}690 \texttt{dm:GeoFeature} footprints, and 316{,}090 \texttt{dm:Image} instances derived from NOAA Emergency Response Imagery, for a total of 5{,}102{,}618 RDF triples. The Policy KG is smaller, containing the \texttt{IoT-Reg} schema, the \texttt{policy-ext} ontology module, and 15 deontic individuals.

Prior to adding imagery, we validated the textual backbone by checking declaration-event linkage, temporal field consistency, and program counts against FEMA documentation. Notably, only 23.4\% of events expose impact-area URLs in source feeds, motivating our separate GeoFeature ingestion.

\paragraph{Multimodal structural validation.}
For the full snapshot, we run SPARQL-based checks targeting cross-modal links: (i) events without locations; (ii) declarations without events; (iii) derived images without \texttt{prov:wasDerivedFrom} links; and (iv) images with inconsistent privacy flags. On the evaluation snapshot, all four metrics are zero: every event has at least one location, every declaration is linked to an event, all derived images carry provenance, and no public image exhibits conflicting flags (Table~\ref{tab:kg-scale-qa}, right), confirming that ingestion pipelines maintain structural invariants before policy testing.

\subsection{Policy Decision Correctness}
\label{sec:eval-decision-correctness}

Sections~\ref{sec:policy-kg} and~\ref{sec:decision_framework} define a deterministic decision logic that fixes the expected verdict and obligations for each relevant combination of data type, audience, and selected flags. We evaluate this by constructing a gold set of packet instances and comparing the system's decisions against expected outcomes.

The gold set consists of 24 packets instantiated from 18 policy scenarios that systematically cover the decision space. The scenarios enumerate all valid combinations of data type (Image, PersonalData), audience (Public, Partner, Internal, Unknown), and compliance state (raw, pre-transformed, error conditions), ensuring that every reachable verdict path is exercised. Core scenarios include (i) imagery released to a public audience with and without prior anonymization; (ii) personal data supplied to a partner agency with and without encryption; (iii) personal data targeted at a public audience (prohibited, regardless of anonymization); and (iv) artifacts directed at audiences lacking explicit permissions (partner for images, internal, unknown). Edge cases include artifacts not present in the DKG, malformed URIs, transform execution failures, over-transformed artifacts satisfying more obligations than required, null compliance flags, re-requests for previously transformed artifacts (idempotency), and unknown data types.
For each packet, we specify the expected initial verdict in \{\textsc{Allow}, \textsc{Block}, \textsc{Allow-with-Transform}\}, the required obligations (if any), and whether a privacy incident should be created. 
Table~\ref{tab:release-decisions} summarizes the 24 gold packets, their expected and observed verdicts, required obligations, and whether a privacy incident was created. Two authors of this paper independently derived the gold-set verdicts from the policy documents and cross-checked them; no disagreements arose because the decision rules in Section~\ref{sec:decision_framework} define a deterministic mapping from requests to verdicts.

\begin{table}[t]
\caption{Policy decision correctness on the gold set of 24 release request packets spanning 18 scenarios, grouped by outcome category.}
\label{tab:release-decisions}
\centering
\setlength{\tabcolsep}{4pt}
\small
\begin{tabular}{@{}llllcc@{}}
\toprule
\textbf{Pkt.} & \textbf{Scenario} & \textbf{Request} & \textbf{Verdict} & \textbf{Xform} & \textbf{Inc.} \\
\midrule
\multicolumn{6}{@{}l}{\textit{Allow (pre-compliant)}} \\
P2 & Pre-anonymized & Img$^\dagger$$\to$Pub & \textsc{Allow} & -- & -- \\
P4 & Pre-encrypted & PII$^\dagger$$\to$Ptnr & \textsc{Allow} & -- & -- \\
P9 & Over-transformed & Img$^\dagger$$\to$Pub & \textsc{Allow} & -- & -- \\
P10 & Over-transformed & PII$^\dagger$$\to$Ptnr & \textsc{Allow} & -- & -- \\
P11 & No PII present & Img$^\circ$$\to$Pub & \textsc{Allow} & -- & -- \\
P15 & Encrypted image & Img$^\dagger$$\to$Ptnr & \textsc{Allow} & -- & -- \\
P23 & Derived artifact & Img$^\prime$$\to$Pub & \textsc{Allow} & -- & -- \\
P24 & Re-request & PII$^\prime$$\to$Ptnr & \textsc{Allow} & -- & -- \\[3pt]
\multicolumn{6}{@{}l}{\textit{Allow via Transform}} \\
P1 & Raw image & Img$\to$Pub & \textsc{AwT}$\to$\textsc{A} & strip & -- \\
P3 & Raw PII & PII$\to$Ptnr & \textsc{AwT}$\to$\textsc{A} & enc & -- \\
P19 & Null flag & Img$\to$Pub & \textsc{AwT}$\to$\textsc{A} & strip & -- \\
P21 & Not retained & Img$\to$Pub & \textsc{AwT}$\to$\textsc{A} & strip & -- \\
P22 & Multi-obligation & PII$\to$Ptnr & \textsc{AwT}$\to$\textsc{A} & enc & -- \\[3pt]
\multicolumn{6}{@{}l}{\textit{Block (prohibited)}} \\
P5 & PII to public & PII$\to$Pub & \textsc{Block} & -- & \cmark \\
P16 & Anon. PII to public & PII$^\dagger$$\to$Pub & \textsc{Block} & -- & \cmark \\[3pt]
\multicolumn{6}{@{}l}{\textit{Block (no permission)}} \\
P6 & No image$\to$partner perm & Img$\to$Ptnr & \textsc{Block} & -- & -- \\
P7 & Unknown audience & PII$\to$Unk & \textsc{Block} & -- & \cmark \\
P14 & Internal audience & PII$\to$Int & \textsc{Block} & -- & \cmark \\
P17 & Unknown audience & Img$\to$Unk & \textsc{Block} & -- & -- \\[3pt]
\multicolumn{6}{@{}l}{\textit{Block (error conditions)}} \\
P8 & Missing file & Img$^\ddagger$$\to$Pub & \textsc{AwT}$\to$\textsc{B} & -- & -- \\
P12 & Not in DKG & PII$^\varnothing$$\to$Ptnr & \textsc{Block} & -- & \cmark \\
P13 & Transform fails & Img$^\times$$\to$Pub & \textsc{AwT}$\to$\textsc{B} & -- & -- \\
P18 & Malformed URI & PII$^\triangle$$\to$Ptnr & \textsc{Block} & -- & \cmark \\
P20 & Unknown type & Sensor$\to$Pub & \textsc{Block} & -- & -- \\
\midrule
\multicolumn{6}{@{}l}{\textbf{Accuracy:} 24/24 (1.0) \quad \textbf{Latency:} $\mu$=0.10\,s, med=0.06\,s, p95=0.18\,s} \\
\bottomrule
\end{tabular}

\vspace{1mm}
\footnotesize
\textsc{AwT}=\textsc{Allow-with-Transform}, \textsc{A}=\textsc{Allow}, \textsc{B}=\textsc{Block}.\\[1pt]
Pub=Public, Ptnr=Partner, Unk=Unknown, Int=Internal.\\[1pt]
$^\dagger$Pre-transformed. $^\circ$No PII. $^\prime$Derived/re-request. $^\ddagger$Missing file. $^\varnothing$Not in DKG. $^\times$Corrupted. $^\triangle$Malformed URI.\\[1pt]
Xform: strip=\texttt{strip\_exif}, enc=\texttt{encrypt\_file}. Inc.=Incident logged.
\end{table}

On this gold set, the system returns the correct verdict and obligations for all 24 cases, yielding exact-match accuracy of~1.0. The median latency per decision is 0.06\,s, with a mean of 0.10\,s and a 95th percentile of 0.18\,s on our single-machine setup. The behavior follows the normative decision logic: for example, non-anonymized imagery to a public audience is assigned \textsc{Allow-with-Transform} with an obligation to obfuscate PII; encrypted personal data to a partner agency is allowed without incident; and personal data addressed to a public audience is blocked and triggers incident creation. Since the policy engine is entirely deterministic and directly implements the specified policy semantics, a perfect score is expected, confirming that the implementation realizes the specified policy semantics on the combinations we intend to support.

\subsection{Transform Impact on Compliance States}
\label{sec:eval-transform-impact}

The previous subsection evaluates whether the system selects the correct verdict and obligations. We also need to verify that, when transforms are required, the resulting artifacts and flags in the DKG are consistent with the intended compliance state. To this end, we run a consistency evaluation over the KG after executing the gold packets, checking three invariants:

\begin{itemize}
    \item \emph{Image anonymization:} for image packets whose initial verdict is \textsc{Allow-with-Transform} with an obfuscation obligation and whose final verdict is \textsc{Allow}, the DKG must contain a derived \texttt{dm:Image} resource that is linked to the original via \texttt{prov:wasDerivedFrom}, marked as anonymized (\texttt{iot-reg:isAnonymized=true}), and no longer marked as containing personal data.
    \item \emph{PII encryption:} for personal-data packets whose initial verdict is \textsc{Allow-with-Transform} with an encryption obligation and whose final verdict is \textsc{Allow}, the DKG must contain a derived artifact that is marked as encrypted (\texttt{iot-reg:isEncrypted=true}) and provenance-linked to the original.
    \item \emph{Incident logging:} for packets whose final verdict is \textsc{Block} and whose data type is \texttt{iot-reg:PersonalData}, the DKG must contain an \texttt{iot-reg:PersonalDataBreach} resource linked to the original artifact and annotated with an incident category and reason.
\end{itemize}

Across the gold packets, all applicable consistency checks passed for image anonymization, PII encryption, and incident logging. Table~\ref{tab:decision-trace} details the artifact state transitions for one such case.

\begin{table}[t]
\centering
\caption{Federated Decision Trace: Partner Agency PII Sharing}
\label{tab:decision-trace}
\small
\begin{tabular}{@{}clll@{}}
\toprule
& \textbf{Graph} & \textbf{Lookup} & \textbf{Result} \\
\midrule
\multicolumn{4}{@{}l@{}}{\textit{Phase 1: Deontic evaluation}} \\[2pt]
1 & PKG & Prohibition match? & None \\
2 & PKG & Permission match & \texttt{:Permit\_PII\_To\_Partner} \\
3 & PKG & Attached obligation & \texttt{:Oblig\_EncryptAndLog} \\
4 & PKG & \texttt{checksFlag} & \texttt{iot-reg:isEncrypted} \\
5 & DKG & Flag satisfied? & \texttt{false} \\
6 & PKG & \texttt{requiresTransform} & \texttt{"encrypt\_file"} \\[2pt]
\midrule
\multicolumn{4}{@{}l@{}}{\textit{Verdict:} \textsc{Allow-with-Transform} $\langle$\texttt{encrypt\_file}$\rangle$} \\
\midrule
\multicolumn{4}{@{}l@{}}{\textit{Phase 2: Transform execution and verification}} \\[2pt]
7 & — & Invoke \texttt{encrypt\_file} & \texttt{dm:Image\_...\_encrypted} \\
8 & DKG & Insert derived artifact & \texttt{isEncrypted} $\leftarrow$ \texttt{true} \\
9 & DKG & Re-verify obligation & Satisfied \\[2pt]
\midrule
\multicolumn{4}{@{}l@{}}{\textit{Final verdict:} \textsc{Allow} — release \texttt{dm:Image\_...\_enc} to \texttt{:PartnerAgency}} \\
\bottomrule
\end{tabular}

\vspace{1mm}
\footnotesize
\noindent \textit{Request:} $r = \langle$\texttt{dm:Image\_17dd9ac6cded\_2005\_Hurricane\_Katrina}, \texttt{:PartnerAgency}, \texttt{iot-reg:DataSharing}, \texttt{iot-reg:PersonalData}$\rangle$.\\[1pt]
\textit{Contrast:} The same artifact with audience \texttt{:PublicAudience} yields \textsc{Block} at step 1 via \texttt{:Prohibit\_Partner\_Reshare}, logging \texttt{dm:Incident\_f1508633...} as an \texttt{iot-reg:PersonalDataBreach}.
\end{table}

\subsection{Query Performance and Scalability}
\label{sec:eval-query}

We now evaluate the behavior of the configured query templates. As described in Section~\ref{sec:monitoring-layer}, the analytical monitoring layer is built on a library of SPARQL query templates over the DKG and PKG, which are exposed to users via a command-line interface. For evaluation, we define two workloads: 21 templates over the DKG alone and 5 federated templates that span both the DKG and PKG.

\paragraph{Single-KG templates.}
The 21 DKG-only templates cover analyst tasks from disaster filtering (by state, year, incident type) to image provenance inspection. Each template exposes a parameterized natural-language prompt (e.g., "show disasters with geofeatures in \{\texttt{state}\}") with a small parameter set, making them reusable across queries without requiring SPARQL expertise. All 21 execute successfully with low latency: see Table~\ref{tab:query-performance}.
\begin{table}[t]
\caption{Latency statistics for the SPARQL query template library, comparing 
single-graph DKG queries against federated DKG+PKG compliance queries 
over the 316K-image evaluation snapshot.}
\label{tab:query-performance}
\centering
\setlength{\tabcolsep}{3.5pt}
\begin{tabular}{@{}lcrrrr@{}}
\toprule
\textbf{Workload} & \textbf{N} & \textbf{Pass} & \textbf{Mean} & \textbf{Med} & \textbf{p95} \\
\midrule
Single-KG (DKG)& 21 & 100\% & 0.05\,s & 0.02\,s & 0.13\,s \\
Federated (DKG+PKG) & 5 & 100\% & 7.0\,s & 5.3\,s & 10.5\,s \\
\bottomrule
\end{tabular}
\end{table}

\paragraph{Federated templates.}
This workload exercises richer policy-aware behavior by issuing federated SPARQL queries that join the DKG and PKG. The five templates implement (i)~a global compliance dashboard summarizing the status of all images; (ii)~a view of images that require transforms before release; (iii)~audience-specific compliance summaries; (iv)~an explanation query that retrieves the policy basis for the decision on a particular image; and (v)~a cross-audience compliance summary for a named event such as Hurricane Katrina. All five templates execute successfully. Latency statistics show a mean of 7.0\,s, a median of 5.3\,s, and a 95th percentile of 10.5\,s. Initial implementations exhibited higher latency due to Cartesian products when joining the 316K-image DKG with PKG rules; reordering \texttt{SERVICE} clauses to query the smaller PKG first and pre-filtering on compliance flags reduced the most expensive query from over 1{,}200\,s to under 5\,s. These interactive latencies compare favorably against manual compliance reviews under FEMA data sharing governance, which can span days to weeks depending on request channel.

The perfect success rates for both workloads follow from design choices: templates are hand-authored and parameterized against the fixed DKG and PKG schemas, and users choose from these templates rather than issuing arbitrary free-form text. Evaluation, therefore, measures whether the configured query library behaves consistently across the breadth of questions it is intended to support, not whether a model can synthesize SPARQL.

\section{Discussion: Generalizability and Limitations}
\label{sec:discussion}

This section discusses the generalizability of our framework to other domains and acknowledges its current limitations.

\subsection{Generalizability}
\label{sec:generalizability}

While our implementation targets FEMA disaster data, the underlying architecture is domain-agnostic. The separation between the domain-specific DKG and the policy-encoding PKG means that adapting the framework to other contexts requires only (i) defining a new domain ontology (analogous to \texttt{dm:}) with artifact types, provenance, and privacy-relevant flags; (ii) instantiating IoT-Reg with domain-appropriate deontic rules, mapping permissions, obligations, and prohibitions to the new artifact types and audiences, and (iii) implementing transform functions specific to the domain's data modalities.

\subsection{Limitations}
\label{sec:limitations}

Our prototype implements \texttt{strip\_exif} for EXIF metadata removal and \texttt{encrypt\_file} for Fernet-based symmetric encryption. Visual PII obfuscation (e.g., face blurring) is not currently implemented; deployments requiring this capability would integrate computer vision pipelines as additional transform functions. However, the framework's design accommodates this: transforms are pluggable, and the verification step in Section~\ref{sec:artifact-derivation} checks compliance flags regardless of how they were set. As such, our evaluation verifies that the system selects the correct verdict and obligations but does not measure the effectiveness of the transforms themselves. Future work could integrate confidence scores from transform pipelines and flag low-confidence transformations for human review, further addressing threat F2 (silent transform failure) from Section \ref{subsec:threat_model}.

The current PKG contains 15 deontic individuals derived from three FEMA/DHS documents. While sufficient for our use case, scaling to hundreds of rules may introduce conflicts not resolvable by simple prohibition dominance. Incorporating defeasible deontic logic or priority annotations is a direction for future work.

Our framework does not model individual consent from disaster survivors. FEMA's operational context relies on statutory authority rather than individual consent, but extending the PKG to incorporate consent-based permissions is future work.

\section{Related Work}
\label{sec:related-work}

We situate our contribution relative to three areas: semantic privacy frameworks, disaster informatics, and policy-aware data systems.

Joshi et al.~\cite{joshi2020integrated} present an integrated knowledge graph for cloud data compliance that captures GDPR and other regulations in a machine-processable format, demonstrating the utility of semantic approaches for privacy compliance automation. LegalRuleML~\cite{governatori2015legalruleml} provides a standard for encoding deontic operators in XML-compatible formats. Our work builds on \texttt{IoT-Reg}~\cite{echenim2023iotreg}, extending it with transform-obligation bindings for disaster imagery. Recent KG-LLM integration for compliance querying~\cite{echenim2025kgllm} complements our deterministic, provenance-verified decision engine.

Pak and Mostafavi~\cite{pak2025situational} argue situational awareness is core to disaster resilience, motivating multi-source data integration. Sanfilippo et al.~\cite{disaster_privacy_2020} analyze the privacy-utility tension in disasters as ``disaster privacy/privacy disaster.'' Our \textsc{Allow-with-Transform} verdict targets this tension directly, enabling conditional sharing under explicit obligations. Prior Disaster KGs focus on event extraction and resource coordination; we integrate privacy compliance as a first-class element.

XACML~\cite{xacml30} provides policy decision points with Permit/Deny/NotApplicable/Indeterminate outcomes and obligation support. Our framework refines this by distinguishing \textsc{Allow} from \textsc{Allow-with-Transform}, binding obligations to transforms and revalidating compliance via provenance-linked derived artifacts before release. Unlike XACML systems that delegate enforcement externally, we embed transform requirements in the policy graph and verify post-transform state against the KG. Purpose-based systems like Hippocratic databases~\cite{agrawal2002hippocratic} track usage against declared purposes. We complement these by making compliance state queryable across linked knowledge graphs.\\

\noindent Our work is distinguished by (1) formal \textsc{Allow-with-Transform} semantics with provenance-verified compliance, (2) application to multimodal disaster artifacts governed by overlapping federal mandates, and (3) end-to-end implementation that spans policy modeling, decision logic, transform execution, and incident logging.

\section{Conclusion}
We have presented a framework that reconciles the tension between operational urgency and privacy compliance in disaster response. By coupling a multimodal Disaster Knowledge Graph with a regulatory Policy Knowledge Graph, we move beyond binary access control to a deontic model supporting \textsc{Allow-with-Transform} decisions. This mechanism binds abstract obligations to executable remediation, such as anonymization or encryption, and verifies compliance through provenance-linked derived artifacts. Evaluation on a 5.1-million-triple snapshot confirms that the system achieves exact-match decision correctness and sub-second latency, providing a viable middle path between data lockdown and unconstrained release. While currently focused on FEMA mandates, this architecture generalizes to any domain requiring policy-compliant sharing of sensitive multimodal data, such as smart cities or healthcare. Future work will explore defeasible reasoning to resolve complex regulatory conflicts and integrate consent-based permissions for survivor-centric privacy.

\begin{acks}
This research was partially supported by a DHS supplement to the NSF award 2310844, IUCRC Phase II UMBC: Center for Accelerated Real time Analytics (CARTA).
\end{acks}


\end{document}